\documentclass[aps, 
amsmath, showpacs, preprintnumbers, twocolumn, amssymb, amsmath, superscriptaddress]{revtex4-1}
\usepackage{graphicx}
\usepackage{mathptmx, textcomp}
\usepackage[latin1]{inputenc}
\usepackage{tabularx}
\usepackage{units}
\usepackage{color}
\usepackage{bm}

\begin{document}
\title{Optical cold damping of neutral nanoparticles near the ground state in an optical lattice}
\author{Mitsuyoshi\,Kamba}
\affiliation{Department of Physics, Tokyo Institute of Technology, 2-12-1 Ookayama, Meguro-ku, 152-8550 Tokyo}
\author{Ryoga\,Shimizu}
\affiliation{Department of Physics, Tokyo Institute of Technology, 2-12-1 Ookayama, Meguro-ku, 152-8550 Tokyo}
\author{Kiyotaka\,Aikawa}
\affiliation{Department of Physics, Tokyo Institute of Technology, 2-12-1 Ookayama, Meguro-ku, 152-8550 Tokyo}

\date{\today}

\pacs{}

\begin{abstract}
We propose and demonstrate purely optical feedback cooling of neutral nanoparticles in an optical lattice to an occupation number of $0.85\pm0.20$. The cooling force is derived from the optical gradients of displaced optical lattices produced with two sidebands on the trapping laser. To achieve highly accurate position observations required for cooling near the ground state, we reduce the laser intensity noise to a relative power noise of $\unit[6\times10^{-8}]{/Hz}$ in a frequency band of $\unit[30]{kHz}$ to $\unit[600]{kHz}$. We establish a reproducible method for neutralizing nanoparticles at high vacuum via a combination of discharging and irradiating an ultraviolet light. Our results form an important basis for the investigation of quantum mechanical properties of ultracold nanoparticles and are also useful for precision measurements with neutral nanoparticles. 
\end{abstract}

\maketitle

\section{Introduction}
\label{sec:intro}

Optically levitated nanoparticles have attracted great interests in recent years~\cite{millen2020optomechanics,gonzalez2021levitodynamics} for their diverse applications, ranging from precision measurements~\cite{moore2014search} and test of fundamental physics~\cite{geraci2010short,bassi2013models,blakemore2021search,aggarwal2022searching,ahn2020ultrasensitive} to various sensory devices such as accelerometers~\cite{monteiro2017optical,hempston2017force,monteiro2020force} and magnetometers~\cite{kumar2017magnetometry}. Extensive studies have realized cooling of their center-of-mass motion to near the ground state~\cite{delic2020cooling,tebbenjohanns2020motional,kamba2021recoil,magrini2021real,tebbenjohanns2021quantum}, opening an intriguing possibility of exploring quantum mechanical properties of the motion of mesoscopic and macroscopic objects~\cite{romero-isart2011large,romero2011optically,bassi2013models}. 

Up to now, three cooling methods have been demonstrated. The first method is parametric feedback cooling (PFC), where the motion of nanoparticles is optically observed and decelerated by modulating the intensity of the trapping laser~\cite{gieseler2012subkelvin,vovrosh2017parametric}.  The second method is cavity cooling, where a high-finesse optical resonator placed near the trapped nanoparticles removes their kinetic energy~\cite{asenbaum2013cavity,kiesel2013cavity,millen2015cavity,delic2020cooling}. The third method is cold damping, where particles are optically observed and their motion is attenuated by an external force synchronized to their motion. In previous studies, the external force has been either radiation pressures~\cite{li2011millikelvin,ranjit2016zeptonewton} or electric forces for charged particles~\cite{iwasaki2019electric,tebbenjohanns2019cold,conangla2019optimal}. While the lowest temperature obtained with PFC is limited to several $\unit[100]{\mu K}$~\cite{jain2016direct}, the other two methods are able to reach temperatures of the order of $\unit[10]{\mu K}$ near the ground state~\cite{delic2020cooling,tebbenjohanns2020motional,kamba2021recoil,magrini2021real,tebbenjohanns2021quantum}. 

The recent progresses in cooling nanoparticles draw attention to the possibility of exploring the quantum mechanical properties of their motion. One of the promising approaches to reveal such properties is the quantum state tomography with time-of-flight measurements~\cite{romero2011optically}, where the momentum of nanoparticles is measured by releasing them from the trapping optical potential. A previous study for time-of-flight measurements with nanoparticles moderately cooled via PFC found that the stray electric fields near surfaces can exert a strong acceleration on nanoparticles~\cite{hebestreit2018sensing}. Stray electric fields near surfaces have been a serious issue also in the field of trapped ions~\cite{brownnutt2015ion}. Such stray fields can be a great obstacle for time-of-flight experiments with nanoparticles cooled to the ground state via cold damping, because only charged nanoparticles have been successfully brought to near the ground state by applying electric feedback forces on them~\cite{tebbenjohanns2020motional,kamba2021recoil,magrini2021real,tebbenjohanns2021quantum}.

Here, we propose and demonstrate a new efficient cooling scheme for neutral nanoparticles, which we call optical cold damping. By applying purely optical forces on neutral nanoparticles in an optical lattice, we realize feedback cooling to an occupation number of $0.85\pm0.20$. We present important advances in optical techniques; first, we enhance the efficiency of collecting the light scattered by nanoparticles with careful alignments of optical layouts. Second, we reduce the noise floor of observing nanoparticles' motion via the active intensity stabilization of the trapping laser to near the shot noise level in a frequency band of $\unit[30]{kHz}$ to $\unit[600]{kHz}$. Laser intensity stabilization near the shot noise limit at such a high frequency range has been relatively unexplored, in spite of numerous previous studies in diverse fields, including gravitational wave detection~\cite{junker2017shot} and cold atom experiments~\cite{wang2020reduction}. Third, we develop an optical setup that allows us to apply controllable optical forces on nanoparticles for cooling their motion. The force originates from the optical gradients of displaced optical lattices that are produced by the weak sidebands on the trapping laser. 

The present work is also an important advance in terms of the wavelength of the laser.The recent studies bringing nanoparticles to the ground state are realized with $\unit[1064]{nm}$ in a room-temperature environment~\cite{delic2020cooling,magrini2021real}, or with $\unit[1550]{nm}$ in a cryogenic environment~\cite{tebbenjohanns2021quantum}. At $\unit[1550]{nm}$, we benefit from the lower absorption of light in nanoparticles and less motional heating via photon scattering than at $\unit[1064]{nm}$. In the present work, despite the difficulty that an optimized objective lens is not readily available at this wavelength, we demonstrate highly accurate observation of the position of nanoparticles, thereby enabling to reach the occupation number of  $0.85\pm0.20$ in a room-temperature environment. 

This article is organized as follows. In Sec.~\ref{sec:theory}, we introduce the theoretical background for understanding of our results. In Sec.~\ref{sec:expsetup}, the details of our experimental setup are explained with a focus on the optical system. In Sec.~\ref{sec:force}, we present the details of the mechanism for the generation of controllable optical forces. In Sec.~\ref{sec:ocd}, we explore the limit of our cooling scheme and show that measurements are in good agreement with the theory of cold damping. Finally, in Sec.~\ref{sec:neutral}, we present a procedure to neutralize nanoparticles at high vacuum via a combined method of a corona discharge and the irradiation of a ultraviolet (UV) light. We summarize our results in Sec.~\ref{sec:concl}

\section{Theoretical description of cold damping}
\label{sec:theory}

 For the comprehensive understanding of our results, we hereby briefly introduce the theoretical description of the motion of nanoparticles in the presence of feedback cooling and various heating mechanisms~\cite{poggio2007feedback,gieseler2012subkelvin,tebbenjohanns2019cold,iwasaki2019electric}.  We ignore heating via the laser phase noise (LPN), which is reduced to a negligibly small value in our setup~\cite{kamba2021recoil}. The one-dimensional equation of motion in the presence of fluctuating forces and damping mechanisms is given by 

\begin{eqnarray}
\label{eq:eom}
\ddot{q} + \Gamma_{\rm tot}\dot{q}+\Gamma_{\rm c}\dot{q_{\rm n}}+\Omega_0^2 q=\dfrac{F_{\rm BG}+F_{\rm r}}{m}
\end{eqnarray}
with $\Gamma_{\rm tot} = \Gamma_{\rm BG}+\Gamma_{\rm c}+\Gamma_{\rm r}$. Here $q$ and $q_n$ denote the position of nanoparticles and the noise in the feedback signal, respectively, while $\Gamma_{\rm BG}$, $\Gamma_{\rm c}$, and $\Gamma_{\rm r}$ denote the damping rate due to collisions with background gases, the damping rate due to feedback cooling, and the damping rate due to photon recoils, respectively. In addition, $\Omega_0$, $m$, $F_{\rm BG}$, and $F_{\rm r}$ denote the oscillation frequency along an optical lattice, the mass of trapped nanoparticles, the stochastic force from background gases, and the stochastic force from photon scattering, respectively.

The effective motional temperature $T_{\rm eff}$ for the particle following Eq.(\ref{eq:eom})  is given as 

\begin{align}
\label{eq:temp}
T_{\rm eff}=&T_0\dfrac{\Gamma_{\rm BG}}{\Gamma_{\rm tot}}+\dfrac{m \Omega_0^2 S_{\rm n}\Gamma_{\rm c}^2}{2k_{\rm B}\Gamma_{\rm tot}}+\dfrac{\hbar \omega_0 P_{\rm sc}}{5m c^2 k_{\rm B}\Gamma_{\rm tot}} \\ \notag
\end{align}
where $T_0$, $S_{\rm n}$, $k_{\rm B}$, $\hbar$, $\omega_0$, $P_{\rm sc}$, $c$ are the temperature of background gases, the power spectral density (PSD) of $q_{\rm n}$, the Boltzmann constant, the reduced Planck constant, the frequency of the trapping light, the optical power scattered by nanoparticles, and the speed of light, respectively. 

At sufficiently low pressures, we can ignore the first term in Eq.(\ref{eq:temp}) and write the occupation number in the presence of feedback cooling  $n_{\rm eq}$ as 

\begin{align}
\label{eq:teff}
n_{\rm eq} +\dfrac{1}{2}=& \dfrac{1}{2\Gamma_{\rm c}\hbar \Omega_0}\left[ 2\gamma_{\rm tot}+m\Omega_0^2S_{\rm n}\Gamma_{\rm c}^2  \right]\\
\gamma_{\rm tot} =& \dfrac{\hbar \omega_0 P_{\rm sc}}{5mc^2}
\end{align}
which provides a minimum value of $n_{\rm eq}+1/2=\sqrt{2\gamma_{\rm tot} m S_{\rm n}}/\hbar$ at an optimum feedback gain of $\Gamma_{\rm c}=\sqrt{2\gamma_{\rm tot}/(m\Omega_0^2S_{\rm n})}$. For our typical experimental parameters, $\Gamma_{\rm c}$ is approximately $\unit[2 \pi \times 5]{kHz}$.

Thus, we find that enhancing the signal-to-noise ratio (SNR) for observing the motion of nanoparticles is crucially important to reach the ground state of the trapping potential. In the context of the theory for a quantum feedback control, it has been known that the efficiency of collecting photons scattered by nanoparticles is directly connected to $n_{\rm eq}$~\cite{clerk2010introduction,doherty2012quantum}.

\section{Experimental setup}
\label{sec:expsetup}

\begin{figure}[t]
\includegraphics[width=0.95\columnwidth] {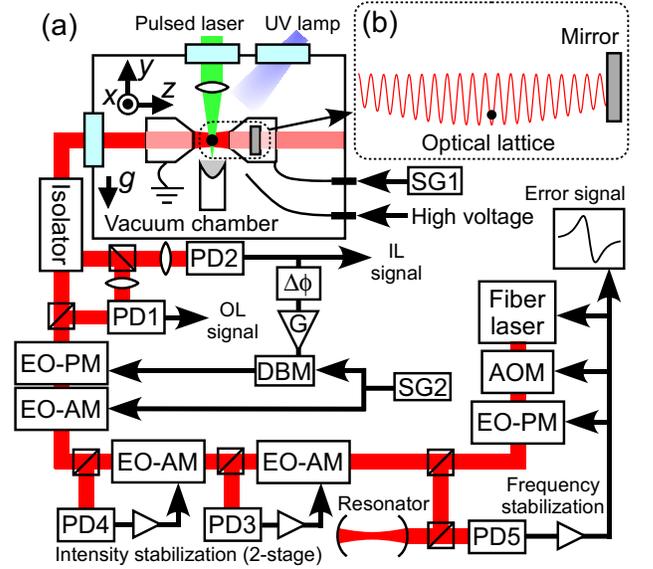}
\caption{(a) Schematic of the experimental setup. PD and SG denote a photodetector and a signal generator, respectively. (b) Single nanoparticles are trapped in an optical lattice formed by retro-reflecting the laser. }
\label{fig:expsetup}
\end{figure}

\subsection{Optical and vacuum systems}
\label{sec:optvac}
Our apparatus mainly consists of an optical system and a vacuum chamber. The schematic of our experimental setup is shown in Fig.~\ref{fig:expsetup}. A single-frequency fiber laser at a wavelength of $2\pi c/\omega_0 = \unit[1550]{nm}$ (NKT Photonics, Koheras C15) with a power of $\unit[140]{mW}$ is incident on a vacuum chamber and is focused by an objective lens with a numerical aperture of 0.85. A one-dimensional optical lattice is formed along the $z$ direction by retro-reflecting the laser with a partially reflective mirror placed inside a vacuum chamber. The distance between the mirror and the trap position $d=\unit[14]{mm}$ is made as short as possible for minimizing the impact of the LPN on the motion of nanoparticles~\cite{kamba2021recoil}. 

We load silica nanoparticles with a radius of $R=\unit[166(3)]{nm}$ and a mass of $\unit[4.4(2) \times 10^{-17}]{kg}$  in an optical lattice at around $\unit[500]{Pa}$ via the pulsed laser deposition of powdery samples placed underneath the trap region~\cite{kamba2021recoil}. We observe nanoparticles with two independent photodetectors; one of which is placed in a feedback loop for cooling and is called in-loop (IL), while the other is used only for estimating $T_{\rm eff}$ in the presence of feedback and is called out-of-loop (OL)~\cite{tebbenjohanns2019cold,conangla2019optimal}. We turn on feedback cooling at around $\unit[10]{Pa}$ and evacuate the chamber to high vacuum. For the charge neutralization of trapped nanoparticles, a deuterium lamp and an electrode for inducing a corona discharge are installed in a vacuum chamber.

In the present study, we focus on the realization of optical cold damping in the $z$ direction, while we cool the motions in the $x$ and $y$ directions via PFC to avoid the escape of trapped nanoparticles at low pressures. We enhance the SNR both by optimizing the optical layouts around the IL and OL photodetectors and by reducing the relative intensity noise (RIN) of the fiber laser. In the following subsections, we explain these crucial aspects of our optical setup. As compared to our previous work~\cite{kamba2021recoil}, where we demonstrated $n_{\rm eq}\sim 3$, these modifications improved the SNR by about a factor of 6, implying cooling to the ground state ($n_{\rm eq}< 1$) is feasible. 

\begin{figure}[t]
\includegraphics[width=0.95\columnwidth] {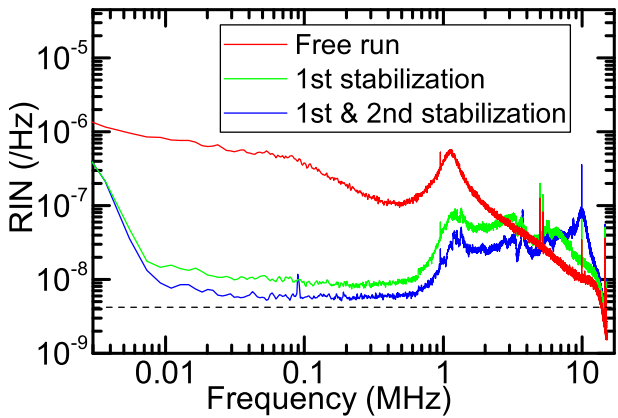}
\caption{RIN as a function of the frequency. The RIN of a free run laser is compared with the RIN with the feedback to the first EO-AM and the RIN with the feedback to both the first and the second EO-AMs. The shot noise level determined by the optical powers on the photodetectors for the stabilization and for the observation is indicated by a dashed line.}
\label{fig:rin}
\end{figure}

\subsection{Enhancement of the signal from nanoparticles}
\label{sec:opt1}

Feedback cooling of nanoparticles to near the ground state requires a high efficiency in collecting scattered photons. In our setup, the scattered light from nanoparticles is collected by the objective lens for focusing the laser and is extracted through an isolator.  Approximately $\unit[85]{\%}$ of the extracted light is incident on the IL photodetector and the remaining light is incident on the OL photodetector. 

The main difference of the present setup from recent studies by other groups with a single-beam optical trap~\cite{delic2020cooling,tebbenjohanns2020motional,magrini2021real,tebbenjohanns2021quantum} is that the scattered light (of the order of $\unit[100]{\mu W}$) is nearly overlapped with the strong retro-reflected beam (about $\unit[40]{mW}$). While we cannot separate these lights, we recognize that their spatial profiles slightly differ with each other, which is presumably due to the complicated spatial profile of the light scattered by nanoparticles~\cite{vamivakas2007phase}. As a result, the amplitude of the signal in the absence of feedback cooling obtained with photodetectors depends on the focal length of the lens placed in front of the photodetectors and the distance between the lens and the photodetectors. We carefully optimized the optical layouts to maximize the amplitude of the signal in the absence of feedback cooling. Incorporating spatial mode-matching of the scattered light and the retro-reflected light in the formalism of ref.~\cite{rahman2018analytical} will allow quantitative understanding of our optimization procedure, which will be a future task.

\subsection{Reduction of the RIN of the laser}
\label{sec:opt2}

The RIN of a commercially available low noise fiber laser is much larger than the PSD of nanoparticles near the ground state. In addition, in our setup, the frequency stabilization introduced for reducing the LPN adds intensity noises at around a few $\unit[100]{kHz}$, close to $\Omega_0/2\pi$.  Therefore, most of previous studies, including ours, have used a balanced detection scheme for subtracting the RIN from the photodetector signals~\cite{gieseler2012subkelvin,iwasaki2019electric}. However, this scheme is ultimately limited by the fact that the subtracting light adds the shot noise on the observation signal and degrades the SNR by $\unit[3]{dB}$, which is a crucial degradation for cooling to the ground state. 

In order to reduce the impact of the RIN on the signal without relying on a balanced detection scheme, we implement the direct reduction of the RIN with electro-optic amplitude modulators (EO-AMs). The advantage of EO-AMs over acousto-optic modulators (AOMs) lies in their large feedback bandwidth of several MHz. In our setup, $\Omega_0/2\pi$ is about $\unit[200]{kHz}$, comparable to the typical bandwidth of AOMs. Therefore, EO-AMs are the only choice in our optical setup. We use two free-space EO-AMs; the first EO-AM plays a dominant role in decreasing the RIN in a frequency range from DC to $\unit[5]{MHz}$, while the second EO-AM is added to further reduce the RIN in a frequency range from $\unit[10]{kHz}$ to $\unit[3]{MHz}$. The feedback signals for the stabilization are generated by home-build PID electronics. In this manner, we obtain the RIN of $\unit[6\times10^{-8}]{/Hz}$, about $\unit[40]{\%}$ above the shot noise level, in a frequency band of $\unit[30]{kHz}$ to $\unit[600]{kHz}$ (Fig.~\ref{fig:rin}). 

With the active intensity stabilization, we achieve a higher value of SNR than with a balanced detection scheme. 
However, the noise floor with the active intensity stabilization is not as flat as the noise floor with a balanced detection scheme, which can be due to two origins. First, the noise floor is currently limited by electronic noises, which are not an ideal white noise. Second, as shown in Fig.~\ref{fig:rin}, the feedback introduces intensity noises at high frequencies of more than $\unit[1]{MHz}$. Unless carefully removed via a low-pass filter, such noises affect the flatness of the noise floor at a low sampling frequency used for acquiring the PSD. 
For the measurement of $T_{\rm eff}$, in particular when the PSD is suppressed to near the noise floor via feedback cooling, it is crucially important to correctly subtract the background level from the integrated PSD. For this reason, we employ a balanced detection scheme for the OL photodetector. Thanks to the flatness of the noise floor, determined purely by the shot noise, we are able to clearly measure the small signal of nanoparticles near the ground state.

\section{Application of an optical cooling force}
\label{sec:force}

\begin{figure}[t]
\includegraphics[width=0.95\columnwidth] {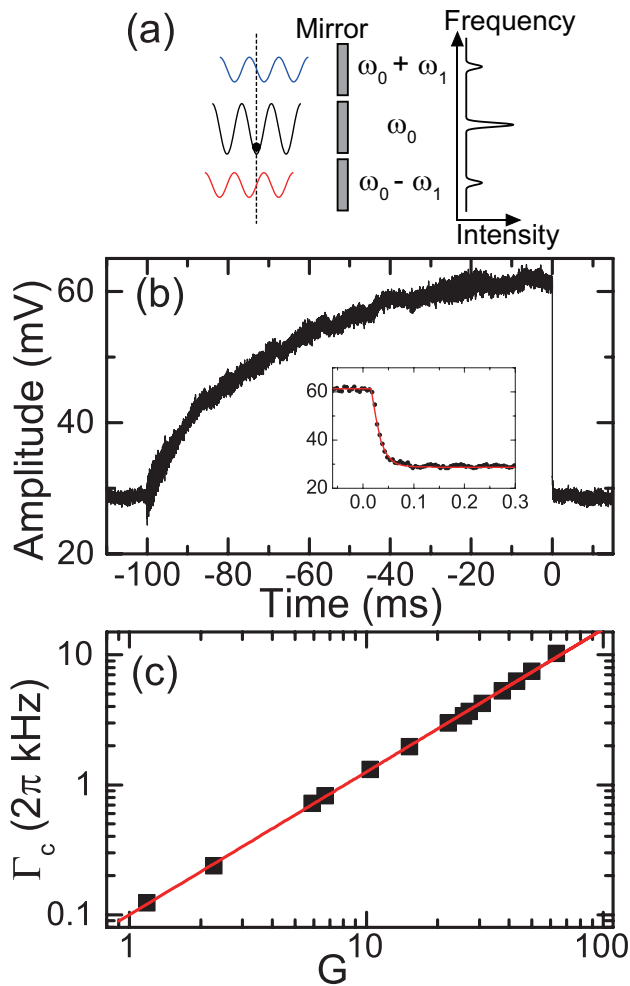}
\caption{(a) Schematic showing the mechanism of exerting optical forces on neutral nanoparticles. Nanoparticles are trapped in an optical lattice with a frequency of $\omega_0$, while the light carries weak sidebands at $\omega_0 \pm \omega_1$ as well. At the position of nanoparticles, the two sidebands exert optical gradient forces with opposite orientations on nanoparticles. By modulating the relative amplitudes of the sidebands, we can apply an oscillatory force on neutral nanoparticles. (b) Time variation of the amplitude of the oscillation signal along the optical lattice at a pressure of $\unit[2\times 10^{-6}]{Pa}$. The trace is averaged over 256 times. When the feedback force is turned off at $t=\unit[-100]{ms}$, the signal increases because of photon recoil heating. When the feedback force is turned on at $t=\unit[0]{ms}$, the oscillation signal rapidly decays as shown in the inset. An exponential fit on the measured decay is shown by a solid line. (c) The damping rate via the cooling force as a function of the feedback gain. The solid line is a linear fit.}
\label{fig:force}
\end{figure}

For the realization of efficient cold damping, it is crucial to apply an external force proportional to the velocity of   nanoparticles. As shown in Section~\ref{sec:theory}, the magnitude of the cooling force has to be larger than a value determined by $\gamma_{\rm tot},m,\Omega_0$ and $S_{\rm n}$. In our previous work with charged nanoparticles, an electric field with a moderate amplitude of a few V, providing a force of the order of $\unit[1\times 10^{-15}]{N}$, was sufficient to reach $n_{\rm eq}\sim 3$~\cite{kamba2021recoil}. We require a new alternative approach that can provide a force of such a magnitude without relying on the charge of nanoparticles. We propose and demonstrate a mechanism based on the gradient of an optical potential.

Figure~\ref{fig:force}(a) shows how our optical potentials exert controllable forces on neutral nanoparticles. We fully take advantage of the standing wave structure that allows us to manipulate trapped nanoparticles via the phase modulation of the laser. We generates weak sidebands (about $\unit[1.4]{\%}$ in amplitude) at frequencies of $\pm \omega_1$ on the trapping laser. While nanoparticles are trapped at the intensity maximum of the optical lattice with $\omega_0$, the optical lattices with $\omega_0\pm \omega_1$ produce gradients at the position of nanoparticles. Thus, the two sidebands exert forces on trapped nanoparticles in the opposite orientations with each other. The magnitudes of these forces depend on the magnitudes of sidebands, implying that an oscillatory cooling force is yielded when the ratio of the magnitudes of the two sidebands are time-varying. We modulate the ratio of the two sidebands with the combination of an EO-AM and an electro-optic phase modulator (EO-PM). The sidebands produced by an EO-AM are in phase with each other, while the sidebands produced by an EO-PM are out of phase with each other. Therefore, the ratio of the magnitudes of the sidebands after an EO-AM and an EO-PM can be controlled by the amplitude of the radio frequency (RF) signal with a frequency of $\omega_1$ applied to the EO-PM. A double balanced mixer (DBM) allows us to control the amplitude and the phase of the RF signal. In this way, we realize the generation of controllable optical forces on trapped nanoparticles with a bandwidth of a few MHz. 

The frequency of the RF has to carefully chosen. For a given amplitude of the sideband, the largest gradient force is obtained at $\omega_1=\pi c/4d$, at which each sideband produces an optical lattice displaced by a quarter of a lattice cite at the trapped position. This frequency corresponds to $2\pi \times \unit[2.7]{GHz}$ in our setup. In the present study, we employ $\omega_1=2\pi\times\unit[450]{MHz}$ because our free-space EO modulators have a bandwidth of a few $\unit[100]{MHz}$. With the present setup, we expect that the magnitude of the optical force is on the order of $\unit[1\times10^{-14}]{N}$, which is larger than we need for realizing cold damping.

To confirm that our idea is properly implemented in our setup, we first measure the amplitude of the cooling force by observing the time variation of the amplitude of the oscillation signal from nanoparticles when the feedback signal is turned on. The feedback signal is obtained by feeding the IL photodetector signal through band-pass filters and an amplifier with a gain of $G$ (Fig.~\ref{fig:expsetup}). Figure~\ref{fig:force}(b) shows an example of such measurements. After we let nanoparticles heated by photon recoils for $\unit[100]{ms}$, we abruptly turn on the feedback signal and observe how fast the signal amplitude decays. The time constant of the decay directly reveals $\Gamma_{\rm c}$ in Eq.(\ref{eq:eom}). Figure ~\ref{fig:force}(c) shows the measured values of $\Gamma_{\rm c}$ for various values of $G$. We find that $\Gamma_{\rm c}$ is proportional to the gain over a wide gain range, suggesting that our modulation scheme is valid in this range and an appropriate feedback signal is generated without any measurable nonlinearity of the DBM. The most important aspect of this measurement is that the maximum value of $\Gamma_{\rm c}$ is turned out to be more than $\unit[2\pi \times 10]{kHz}$, which is larger than the value we need for cooling nanoparticles to near the ground state~\cite{kamba2021recoil,tebbenjohanns2021quantum}. Thus, we find that our scheme is a promising method for cold damping of neutral nanoparticles to near the ground state.

\section{Optical cold damping}
\label{sec:ocd}

\begin{figure}[t]
\includegraphics[width=0.95\columnwidth] {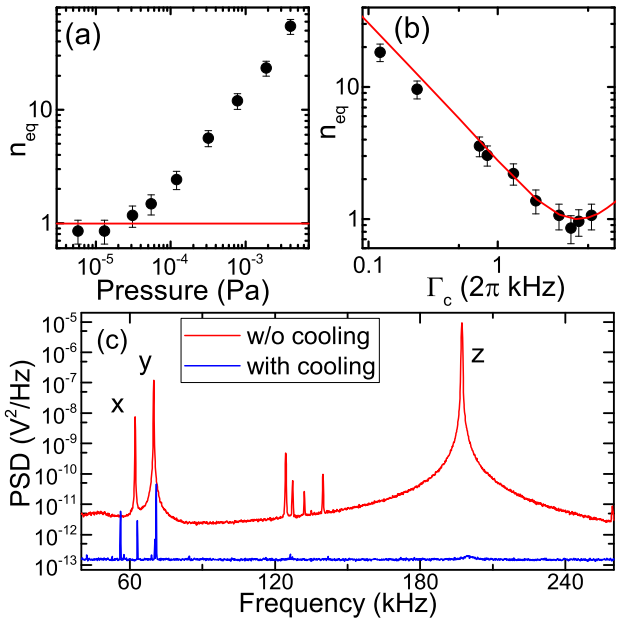}
\caption{(a) The occupation number with respect to the pressure for $\Gamma_{\rm c} = \unit[2\pi \times 3.7]{kHz}$. The error bars indicate systematic thermal fluctuations in measuring $T_{\rm eff}$. The calculated value at low pressures obtained with Eq.(\ref{eq:teff}) is shown by the solid line. (b) The occupation number with respect to $\Gamma_{\rm c}$ at a pressure of $\unit[5\times 10^{-6}]{Pa}$. The error bars indicate systematic thermal fluctuations in measuring $T_{\rm eff}$.  The solid line is not a fit and shows the calculated values with Eq.(\ref{eq:teff}) for our experimental parameters. (c) PSDs of the OL signal with and without feedback cooling obtained at $\unit[6\times10^{-6}]{Pa}$ and $\unit[6]{Pa}$, respectively. }
\label{fig:ocd}
\end{figure}

We investigate the limit of our new cooling approach by measuring $n_{\rm eq}$ for various experimental conditions. In these measurements, we first measure $T_{\rm eff}$ by comparing the PSDs with and without feedback cooling [Fig.~\ref{fig:ocd}(c)]~\cite{gieseler2012subkelvin,vovrosh2017parametric}, after which $n_{\rm eq}$ is calculated from $T_{\rm eff}$. Figure~\ref{fig:ocd}(a) shows the measured $n_{\rm eq}$ with respect to the pressure, while Figure~\ref{fig:ocd}(b) shows the measured $n_{\rm eq}$ with respect to $\Gamma_{\rm c} $. At pressures lower than $\unit[2\times 10^{-5}]{Pa}$, we find that the observed $n_{\rm eq}$ agrees well with the theoretically expected value of Eq.(\ref{eq:teff}) with our experimental parameters. The calculation with Eq.(\ref{eq:teff}) is also in good agreement with the measurements of Fig.~\ref{fig:ocd}(b) without any fitting parameter. The lowest $T_{\rm eff}$ and the lowest $n_{\rm eq}$ are $T_{\rm eff} \sim \unit[13 \pm 2]{\mu K}$ and $n_{\rm eq} \sim 0.85\pm0.20$, respectively, where the error indicates systematic errors from thermal fluctuations. 

For comparison, we test conventional electric cold damping for charged nanoparticles in the present setup, which is readily realized by applying the feedback signal to the electrode along the $z$ direction (the right lens in Fig.~\ref{fig:expsetup}). We find that the obtained $n_{\rm eq}$ is comparable to that with optical cold damping. Furthermore, we find that optical cold damping is even more stable than electric cold damping. Upon decreasing the pressure, we often observe that the charge number varies at pressures between $1$ and $\unit[1\times 10^{-4}]{Pa}$ and, in some cases, the charge is inverted, resulting in the loss of nanoparticles under electric cold damping. Even if they are not lost, with electric cold damping, the feedback gain has to be carefully adjusted because the ratio of the charge to the mass depends on nanoparticles and can vary during experiments. Such issues are of never concern with optical cold damping. The amplitude of the optical gradient force on nanoparticles is proportional to their polarizability, which is proportional to $m$~\cite{gieseler2012subkelvin}, indicating that the acceleration of nanoparticles caused by a given amplitude of the feedback signal is independent from $m$. Therefore, fixed experimental parameters can be used for various nanoparticles. It is obvious that, although our scheme is developed for cooling neutral nanoparticles, it works similarly well for charged nanoparticles.

\section{Charge neutralization}
\label{sec:neutral}

\begin{figure}[t]
\includegraphics[width=0.95\columnwidth] {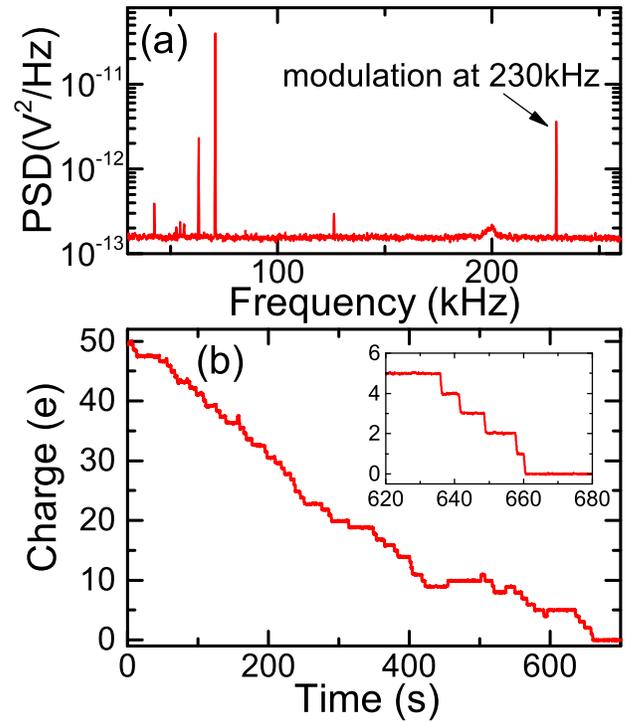}
\caption{(a) PSD of singly charged nanoparticles with a sinusoidal modulation via an external electric field at a pressure of  $\unit[4\times10^{-5}]{Pa}$. The peak at $\unit[230]{kHz}$ arises due to the modulation. (b) Typical time variation of the charge of a trapped nanoparticle under the illumination of a UV lamp at around $\unit[4\times10^{-5}]{Pa}$. In the inset, an expanded plot near the moment of neutralization is shown. }
\label{fig:neutral}
\end{figure}

Charge neutralization has been an imperative issue in diverse fields, including gravitational wave detection~\cite{armano2018precision}, inertial sensors~\cite{ciani2017new}, and precision measurements with levitated microparticles~\cite{moore2014search,monteiro2020force,blakemore2021search}. In many of these studies, irradiating UV lights has been a powerful means for neutralization. In a recent work with nanoparticles, a corona discharge with a voltage of several kV was employed for the charge management~\cite{frimmer2017controlling}. In our setup, 
we find that the charge number of nanoparticles often varies at pressures between $1$ and $\unit[1\times10^{-4}]{Pa}$. In particular, a frequent charge variation is observed when nanoparticles are brought to low pressures for the first time after they are loaded into an optical lattice. In many cases, nanoparticles are negatively charged after the evacuation. We infer that this behavior is relevant to outgassing from the surface of nanoparticles. Therefore, charge neutralization has to be carried out at pressures of $<\unit[1\times 10^{-4}]{Pa}$. 

We tested both methods, discharging and irradiating a UV light, in our setup and found that neither of which is suitable for our application. On the one hand, a UV light from a deuterium lamp is found to be strong enough for neutralizing positively charged nanoparticles even at high vacuum, while it turns out that it cannot neutralize negatively charged nanoparticles because it provides dominantly negative charges. On the other hand, although a corona discharge readily occurs at a voltage of $\unit[500]{V}$ at low to medium vacuum and can neutralize both positively and negatively charged nanoparticles, a discharge hardly occurs with our high voltage capability at high vacuum. 

We find that a combined method is the easiest and the most reproducible way to neutralize nanoparticles and to keep them neutralized for many hours. We first apply a high voltage at around $\unit[350]{Pa}$, where nanoparticles are positively charged with a typical charge of about $50e$ with $e$ being the elementary charge. The charge number is measured by observing the response of nanoparticles to a sinusoidal electric field oscillating at a frequency close to $\Omega_0/2\pi$ [Fig.~\ref{fig:neutral}(a)]. After we confirm that nanoparticles are positively charged, we evacuate the chamber to below $\unit[1\times 10^{-4}]{Pa}$ and apply a UV light. Typically nanoparticles are neutralized in several minutes [Fig.~\ref{fig:neutral}(b)]. Nanoparticles neutralized in this manner stay neutralized for tens of hours, although rarely they get charged after several hours. In such a case, we repeat the procedure stated above. 

\section{Conclusion}
\label{sec:concl}

We have developed a purely optical feedback cooling scheme for neutral nanoparticles in an optical lattice and demonstrated cooling their motion to an occupation number of $0.85\pm0.20$. For this purpose, we improved the SNR of observing the motion of nanoparticles both by enhancing the efficiency of collecting the light scattered by nanoparticles and by lowering the RIN of the laser to near the shot noise in a frequency band of $\unit[30]{kHz}$ to $\unit[600]{kHz}$. The strong optical force required for efficient cooling is derived from the optical gradients of the displaced optical lattices generated by two weak sidebands on the trapping laser. By modulating the relative amplitudes of the two sidebands, we exert oscillatory cooling forces on trapped nanoparticles. In addition to realizing the new cooling approach, we established a reproducible procedure to neutralize nanoparticles suitable for experiments at high vacuum.

Our scheme is superior to PFC, the most commonly used optical cooling approach, in terms of the following aspects. First, our scheme modulates the laser at frequencies of a few $\unit[100]{MHz}$, much higher than typical oscillation frequencies of nanoparticles. This is in contrast to PFC, which introduces an intensity noise at twice the oscillation frequency with an amplitude higher than that of the shot noise by orders of magnitude. Second, due to the strong damping provided by our scheme, the lowest $n_{\rm eq}$ achieved with our scheme is about two orders of magnitude lower than with PFC~\cite{jain2016direct}. 

In comparison with cavity cooling, which can also be a powerful approach for neutral nanoparticles, our scheme provides comparable $n_{\rm eq}$, while the high accuracy of observing the motion of nanoparticles, about $\unit[10]{fm/\sqrt{Hz}}$, which is naturally provided by the optics for feedback cooling, is strongly beneficial for the application in sensitive accelerometers. Furthermore, the simplicity of our scheme that the trapping laser also carries the mechanism of cooling may facilitate the momentum spectroscopy via time-of-flight. With our scheme, time-of-flight measurements are readily implemented just by turning off the trapping laser. Our results also form the important basis for applications in precision measurements and accelerometers, where ultracold neutral particles are required~\cite{monteiro2020force,blakemore2021search}.

In the present work, we demonstrate optical cold damping near the ground state only in one direction, while the motions in other directions are cooled via PFC and their temperatures are still high. For future experiments, it is desirable to cool the motions in all directions to near the ground state, which can be realized by producing controllable optical gradients perpendicular to the light propagation directions either via a two-dimensional AOM, or via three dimensional optical lattices. 

The present work represents the unique possibility of manipulating nanoparticles in an optical lattice via the phase modulation of the trapping laser, which can be potentially useful for accelerating microscopic particles such as atoms and molecules. We envision that the phase modulation technique will open further exciting possibilities such as the generation of anharmonic potentials~\cite{schmelcher2018driven,dhar2019run} and the exploration of the physics with a time-dependent optical lattice~\cite{lignier2007dynamical,hauke2012non,ha2015roton}.

\textit{Note added:} After the submission of the present work, a related work on an all-optical approach for cold damping appeared on arXiv~\cite{vijayan2022scalable}.

\begin{acknowledgments}
We thank M.\,Kozuma and H.\,Kanamori for fruitful discussions. We are grateful to S.\,Nakano and N.\,Kagatani for their experimental assistance. This work is supported by the Murata Science Foundation, the Mitsubishi Foundation, the Challenging Research Award, the 'Planting Seeds for Research' program, and STAR Grant funded by the Tokyo Tech Fund, Research Foundation for Opto-Science and Technology, JSPS KAKENHI (Grants No. JP16K13857, JP16H06016, and JP19H01822), JST PRESTO (Grant No. JPMJPR1661), and COI-NEXT (Grant No. JPMJPF2015).
\end{acknowledgments}

\bibliographystyle{apsrev}


\end{document}